\documentclass[a4paper]{mem}
\usepackage{natbib}
\usepackage{graphicx}
\usepackage[a4paper]{hyperref}
\idline{00}{0}

\def\Reff{R_{\mathrm e}}

\def\csiv{{\vec{x}'}}
\def\nuv{{\vec{v}'}}
\def\Omegv{\vec{\Omega}}
\def\xv{\vec{x}}

\def\csici{x'_i}

\def\ddcsiv{\ddot{\csiv}}

\def\dOmegv{\dot{\Omegv}}
\def\ddxv{\ddot{\xv}}

\def\Rot{{\cal R}}
\def\RotT{\Rot^{\rm T}}

\def\phiM{\varphi_{\rm M}}
\def\theM{\vartheta_{\rm M}}
\def\psiM{\psi_{\rm M}}

\def\omegphi{\omega_{\varphi}}
\def\omegthe{\omega_{\vartheta}}
\def\omegpsi{\omega_{\psi}}

\def\dphi{\dot\varphi}
\def\dtheta{\dot\vartheta}
\def\dpsi{\dot\psi}

\def\ddphi{\ddot\varphi}
\def\ddtheta{\ddot\vartheta}
\def\ddpsi{\ddot\psi}

\def\Tc{\vec{T}}
\def\Ti{T_i}
\def\Tj{T_j}
\def\Tu{T_1}
\def\Td{T_2}
\def\Tt{T_3}

\def\DT{\Delta T}
\def\DTij{\DT_{ij}}
\def\DTdu{\DT_{21}}
\def\DTtu{\DT_{31}}
\def\DTtd{\DT_{32}}
\def\DTud{\DT_{12}}
\def\DTut{\DT_{13}}
\def\DTdt{\DT_{23}}

\def\Ii{I_i}
\def\Iu{I_1}
\def\Id{I_2}
\def\It{I_3}
\def\DI{\Delta I}
\def\DIdu{\DI_{21}}
\def\DItu{\DI_{31}}
\def\DItd{\DI_{32}}

\def\phig{\phi_{\mathrm g}}
\def\rhog{\rho_{\mathrm g}}
\def\Mg{M_{\mathrm g}}
\def\Mgn{M_{\mathrm g,11}}
\def\hg{h_{\mathrm g}}

\def\alphau{\alpha_1}
\def\alphad{\alpha_2}
\def\alphat{\alpha_3}
\def\alphai{\alpha_i}
\def\alphaj{\alpha_j}
\def\alphak{\alpha_k}
\def\alphaun{\alpha_{1,1}}

\def\acu{a_1}
\def\acd{a_2}
\def\act{a_3}

\def\acun{a_{1,250}}

\def\Porb{P_{\mathrm orb}}
\def\Pdyn{P_{\mathrm dyn}}

\def\msun{M_{\odot}}

\def\rhoc{\rho_{\mathrm c}}
\def\rhocz{\rho_{\mathrm c,0}}
\def\Mc{M_{\mathrm c}}

\def\sigv{\sigma_{\mathrm V}}
\def\sigvn{\sigma_{\mathrm V,1000}}
\def\sigvd{\sigv^2}

\def\Pphi{P_{\varphi}}
\def\Pthe{P_{\vartheta}}
\def\Ppsi{P_{\psi}}

\def\tH{t_{\mathrm H}}

\def\wci{w_i}
\def\wcu{w_1}
\def\wcd{w_2}
\def\wct{w_3}

\def\Ntot{N_{\rm tot}}
\def\Nesc{N_{\rm esc}}

\begin{document}
   \title{Elliptical galaxies interacting with the cluster
          tidal field: origin of the intracluster stellar population}

   \author{V. Muccione \inst{1}, L. Ciotti\inst{2}}

   \institute{Geneva Observatory,51 ch. des Maillettes, 
              1290 Sauverny, Switzerland\\ 
              \email{veruska.muccione@obs.unige.ch}\\ 
              \and Dipartimento di Astronomia, Universit\`a di Bologna, 
               via Ranzani 1, 40127 Bologna, Italy\
              \email{ciotti@bo.astro.it}}

\abstract{

With the aid of simple numerical models, we discuss a particular
aspect of the interaction between stellar orbital periods inside
elliptical galaxies (Es) and the parent cluster tidal field (CTF),
i.e., the possibility that {\it collisionless stellar evaporation}
from Es is an effective mechanism for the production of the recently
discovered intracluster stellar populations (ISP). These very
preliminary investigations, based on idealized galaxy density profiles
(such as Ferrers density distributions) show that, over an Hubble
time, the amount of stars lost by a representative galaxy may sum up
to the 10\% of the initial galaxy mass, a fraction in interesting
agreement with observational data.  The effectiveness of this
mechanism is due to the fact that the galaxy oscillation periods near
its equilibrium configurations in the CTF are of the same order of
stellar orbital times in the external galaxy regions.
   
   }
   \authorrunning{V. Muccione \& L. Ciotti}
   \titlerunning{Origin of the ISP}
   \maketitle
%

\section{Introduction}

Observational evidences of an Intracluster Stellar Population (ISP)
are mainly based on the identification of {\it intergalactic}
planetary nebulae and red giant branch stars (see, e.g., Theuns \&
Warren 1996, M\`endez et al. 1998, Feldmeier et al. 1998, Arnaboldi et
al. 2002, Durrell et al. 2002).  Overall, the data suggest that
approximately 10\% (or even more) of the stellar mass of the cluster
is contributed by the ISP (see, e.g., Ferguson \& Tanvir 1998).

The usual scenario assumed to explain the finding above is that
gravitational interactions between galaxies in cluster and/or
interactions between the galaxies with the tidal gravitational field
of the parent cluster (CTF), lead to a substantial stripping of stars
from the galaxies themselves.

Here, supported by a curious coincidence, namely by the fact that {\it
the characteristic times of oscillation of a galaxy around its
equilibrium position in the CTF are of the same order of magnitude of
the stellar orbital periods in the external part of the galaxy
itself}, we suggest that an additional effect is at work, i.e. we
discuss about the possible ``resonant'' interaction between stellar
orbits inside the galaxies and the CTF.

In fact, based on the observational evidence that the major axis of
cluster Es seems to be preferentially oriented toward the cluster
center, N-body simulations showed that galaxies tend to align reacting
to the CTF as rigid (Ciotti \& Dutta 1994).  By assuming this
idealized scenario, a stability analysis then shows that this
configuration is of equilibrium, and allows to calculate the
oscillation periods in the linearized regime (Ciotti \& Giampieri
1998, hereafter CG98).

Prompted by these observational and theoretical considerations, in our
numerical explorations we assume that the galaxy is a triaxial
ellipsoid with its center of mass at rest at the center of a triaxial
cluster. The considerably more complicate case of a galaxy with the
center of mass in rotation around the center of a spherical cluster
will be discussed elsewhere (Ciotti \& Muccione 2003, hereafter
CM03).

\section{The physical background}

Without loss of generality we assume that in the (inertial) Cartesian
coordinate system $C$ centered on the cluster center, the CTF tensor
$\Tc$ is in diagonal form, with components $\Ti$ $(i=1,2,3)$.  By
using three successive, counterclockwise rotations ($\varphi$ around
$x$ axis, $\vartheta$ around $y'$ axis and $\psi$ around $z''$ axis),
the linearized equations of motion for the galaxy near the equilibrium
configurations can be written as
\begin{eqnarray}
\ddphi={\DTtd\DItd\over\Iu}\varphi, \nonumber \\ 
\ddtheta={\DTtu\DItu\over\Id}\vartheta, \nonumber \\
\ddpsi={\DTdu\DIdu\over\It}\psi,
\label{eqn1}
\end{eqnarray}
where $\Delta T$ is the {\it antisymmetric} tensor of components
$\DTij \equiv \Ti-\Tj$, and $\Ii$ are the principal components of the
galaxy inertia tensor.  In addition, let us also assume that
$\Tu\geq\Td\geq\Tt$ and $\Iu\leq\Id\leq\It$, i.e., that $\DTtd, \DTtu$
and $\DTdu$ are all less or equal to zero (see Section 3). Thus, the
equilibrium position of \ref{eqn1} is {\it linearly stable}, and its
solution is
\begin{eqnarray}
\varphi   =\phiM \cos (\omegphi t), \nonumber \\ 
\vartheta =\theM \cos (\omegthe t), \nonumber \\
\psi      =\psiM \cos (\omegpsi t),
\label{eqn2}
\end{eqnarray}
where
\begin{eqnarray}
\omegphi = \sqrt{\DTdt\DItd\over\Iu},\nonumber \\  
\omegthe = \sqrt{\DTut\DItu\over\Id}, \nonumber \\
\omegpsi = \sqrt{\DTud\DIdu\over\It}.
\label{eqn3}
\end{eqnarray}

For computational reasons the best reference system in which calculate
stellar orbits is the (non inertial) reference system
$C'$ in which the galaxy is at rest, and its inertia tensor is in
diagonal form. The equation of the motion for a star in $C'$ is
\begin{equation}
\ddcsiv = \Rot^{\rm T}\ddxv 
        -2\Omegv\wedge\nuv 
        -\dOmegv\wedge\csiv 
        -\Omegv\wedge (\Omegv\wedge\csiv ),
\label{eqn4}
\end{equation}
where $\xv=\Rot (\varphi,\vartheta,\psi)\csiv$, and
\begin{eqnarray} 
\Omegv &=& (\dphi\cos\vartheta\cos\psi +\dtheta\sin\psi , 
           -\dphi\cos\vartheta\sin\psi \nonumber \\
       & &  +\dtheta\cos\psi,\dphi\sin\vartheta +\dpsi).
\label{eqn5}
\end{eqnarray}
In eq. (4)
\begin{equation}
\RotT\ddxv = -\nabla_{\csiv}\phig +(\RotT\Tc\Rot)\csiv,
\label{eqn6}
\end{equation}
where $\phig (\csiv)$ is the galactic gravitational potential,
$\nabla_{\csiv}$ is the gradient operator in $C'$, and we used the
tidal approximation to obtain the star acceleration due to the cluster
gravitational field. 

\section{Period estimations}

For simplicity, in the following estimates we assume that the galaxy
and cluster densities are stratified on homeoids.  In particular, we
use a galaxy density profile belongings to the ellipsoidal
generalization of the widely used $\gamma$-models (Dehnen 1993,
Tremaine et al. 1994):
\begin{equation}
\rhog (m) = {\Mg\over \alphau\alphad\alphat}
{3-\gamma\over 4\pi} {1\over m^\gamma (1+m)^{4-\gamma}},
\label{eqn7}
\end{equation} 
where $\Mg$ is the total mass of the galaxy, $0\leq\gamma\leq 3$, and
\begin{equation}
m^2 =  \sum_{i=1}^3 {(\csici )^2\over\alphai^2},
\qquad \alphau\geq\alphad\geq\alphat.
\label{eqn8}
\end{equation}

The inertia tensor components for this family are given by
\begin{equation}
\Ii={4\pi\over 3}\alphau\alphad\alphat (\alphaj^2+\alphak^2)\hg,
\label{eqn9}
\end{equation}
where $\hg =\int_0^{\infty}\rhog (m)m^4 dm$, and so
$\Iu\leq\Id\leq\It$.  Note that, from eq. (3), {\it the frequencies for
homeoidal stratifications do not depend on the specific density
distribution assumed, but only on the quantities $(\alphau ,\alphad
,\alphat)$}. We also introduce the two ellipticities
\begin{equation}
{\alphad\over\alphau}\equiv 1 -\epsilon, \quad\quad
{\alphat\over\alphau}\equiv 1 -\eta,
\label{eqn10}
\end{equation}
where $\epsilon\leq\eta\leq 0.7$ in order to reproduce realistic
flattenings.

A rough estimate of {\it characteristic stellar orbital times} inside
$m$ is given by $\Porb (m)\simeq 4\Pdyn (m) = \sqrt{3\pi
/G\overline{\rhog} (m)}$, where $\overline{\rhog}(m)$ is the mean
galaxy density inside $m$. We thus obtain 
\begin{eqnarray}
\Porb (m)&\simeq& 9.35\times 10^6 
                  \sqrt{{\alphaun^3 (1-\epsilon)(1-\eta)\over\Mgn}}
                  \nonumber \\
         &      & \times m^{\gamma/2}(1+m)^{({3-\gamma})/2} 
                  \;{\mathrm{yrs}}, 
\label{eqn11}
\end{eqnarray}
where $\Mgn$ is the galaxy mass normalized to $10^{11}\msun$,
$\alphaun$ is the galaxy ``core'' major axis in kpc units (for the
spherically symmetric $\gamma=1$ Hernquist 1990 models, $\Reff\simeq
1.8 \alphau$); thus, in the outskirts of normal galaxies orbital times
well exceeds $10^8$ or even $10^9$ yrs.

For the cluster density profile we assume
\begin{equation}
\rhoc (m) = {\rhocz\over (1+m^2)^2},
\label{eqn12}
\end{equation}    
where $m$ is given by an identity similar to eq. (8), with
$\acu\geq\acd\geq\act$, and, in analogy with eq. (10), we define
$\acd/\acu\equiv 1-\mu$ and $\act/\acu\equiv 1-\nu$, with
$\mu\leq\nu\leq 1$.

It can be shown that the CTF components at the center of a
non-singular homeoidal distribution are given by
\begin{equation}
\Ti=-2\pi G\rhocz\wci (\mu,\nu),
\label{eqn13}
\end{equation}
where the dimensionless quantities $\wci$ are independent of the
specific density profile, $\wcu\leq\wcd\leq\wct$ for
$\acu\geq\acd\geq\act$, and so the conditions for stable equilibrium in
eq. (1) are fulfilled (CG98, CM03).

\begin{figure*}
   \centering
   \resizebox{\hsize}{!}{\includegraphics[clip=true]{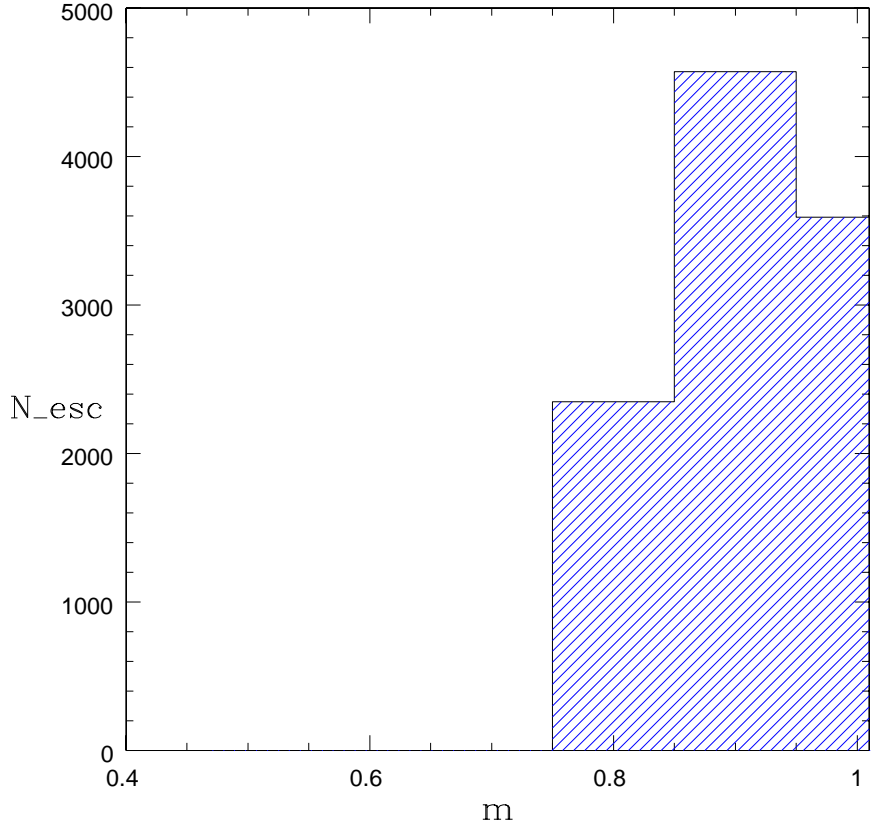}}
     \caption{Histogram of $m$ at $t=0$ vs. $\Nesc$ for a galaxy 
              of $\Mg=10^{10}\,\msun$ and $\alphau =20$ kpc, with 
              $\epsilon =0.2$ and $\eta=0.4$. The cluster parameters 
              are $\mu=0.2$, $\nu =0.4$, $\acun= \sigvn =1$, and the 
              galaxy maximum oscillations angles in eqs. (2) are fixed 
              to 0.2 rad. In this simulation $\Ntot=10^5$.}  
        \label{fig1}
    \end{figure*}

The quantity $\rhocz$ is not a well measured quantity in real
clusters, and for its determination we use the virial theorem,
$\Mc\sigvd = -U$, where $\sigvd$ is the virial velocity dispersion,
that we assume to be estimated by the observed velocity dispersion of
galaxies in the cluster. Thus, we can now compare the galactic
oscillation periods (for small galaxy and cluster flattenings, see
CM03)
\begin{eqnarray}
\Pphi &=& {2\pi\over \omegphi} 
          \simeq {8.58\times 10^8\over\sqrt{(\nu -\mu)(\eta -\varepsilon)}}
          {\acun\over\sigvn} \; \mathrm{yrs} \; , \nonumber \\
\Pthe &=& {2\pi\over \omegthe}
          \simeq {8.58\times 10^8\over\sqrt{\nu\eta}}
          {\acun\over\sigvn} \; \mathrm{yrs}\; , \nonumber \\
\Ppsi &=& {2\pi\over\omegpsi} 
          \simeq {8.58\times 10^8\over\sqrt{\mu\varepsilon}}
          {\acun\over\sigvn} \; \mathrm{yrs}\; .
\label{eqn14}
\end{eqnarray}
(where $\acun = \acu/250$kpc, and $\sigvn =\sigv/10^3$km s$^{-1}$)
with the characteristic orbital times in galaxies. From eqs. (11) and
(14), it follows that {\it in the outer halo of giant Es the stellar
orbital times can be of the same order of magnitude as the oscillatory
periods of the galaxies themselves in the CTF}.  For example, in a
relatively small galaxy of $\Mgn =0.1$ and $\alphaun =1$, $\Porb\simeq
1$ Gyr is at $m\simeq 10$ (i.e., at $\simeq 5\Reff$), while for a
galaxy with $\Mgn =1$ and $\alphaun =3$ the same orbital time is at
$m\simeq 7$ (i.e., at $\simeq 3.5\Reff$).

\section{The numerical integration scheme}

Here we present the preliminary results obtained by using only a very
idealized galactic density profile, while more realistic galaxy
density distributions, such as triaxial Hernquist models, are
described elsewhere (CM03, Muccione \& Ciotti 2003, hereafter
MC03). In particular, here we explore the evaporation process in the
case of a Ferrers (see, e.g., Binney \& Tremaine 1987) models, with
density profiles given by
\begin{eqnarray}
\rho_g & = & 
\left\{
\begin{array}{lr}
\rho_g(0) (1-m^2)^n \,\, & \mbox{for}\, m \leq 1 \\
0 & \mbox{for}\, m > 1 
\end{array}
\right.
\label{eqn15}
\end{eqnarray}
where $m$ is the homeoidal radius defined as in eq. (8).  The case
$n=0$ corresponds to an anisotropic harmonic oscillator, while for
$n\geq 0$ the density distribution is radially decreasing. A nice
property of Ferrers models (for integer $n$) is that their
gravitational potential can be simply expressed in algebraic form
(see, e.g., Chandrasekhar 1969).

To integrate the second order differential equations (11) for each
star, we use a code based on an adapted Runge-Kutta routine.  The
initial conditions are obtained by using the Von Neumann {\it
rejection method}, and with this method we arrange $\Ntot$ initial
conditions in configuration space which reproduce the density profile
in eq. (15). In MC03 and CM03 the three components of the initial
velocity for each star are assigned by considering the local velocity
dispersion of the galaxy model at rest, while here, for simplicity,
each star at $t=0$ is characterized by null velocity.  Within $t=\tH$
(where $\tH =1.5\times 10^{10}$ yrs), the code checks which of the
initial conditions result in a orbit with $m >1$. This escape
condition is very crude, and a more sophisticated criterium is adopted
in MC03 and CM03, where {\it untruncated} galaxy models are used. In
standard simulations we use, as a rule, $\Ntot =10^5$, thus, from this
point of view, we are exploring $10^5$ independent ``1-body problems''
in a time--dependent force field.

The simulations are performed on the {\it GRAVITOR}, the Geneva
Observatory 132 processors Beowulf cluster ({\bf http://obswww.
unige.ch/\~{}pfennige/gravitor/ \\ gravitor\_e.html}).

\section{Preliminary results and conclusions}
 
In Fig. 1 we show the results of one of our preliminary simulations
for a galaxy model with $n=1$, $\Mg =10^{10}\,\msun$, semi-mayor axis
$\alphau =20$ kpc, flattenings $\epsilon =0.2$, $\eta=0.4$, and
maximum oscillation angles equals to 0.2 rad. The cluster parameters
are $\acun=\sigvn =1$, $\mu=0.2$, $\nu=0.4$, and the total number of
explored orbits is $\Ntot=10^5$. In the ordinate axis we plot the
number of {\it escapers} as a function of their homeoidal radius at
$t=0$.  Note how the zone of maximum escape is near $m\simeq 0.8$, and
how the total number of escapers is a significant fraction of the
total number of explored orbits (actually, with the galaxy and cluster
parameters adopted in the simulations, $N_{\rm esc}\simeq 0.1\Ntot$).
This number is a somewhat upper limit of the expected number in more
realistic simulations: in fact, 1) as described in Section 4, we
adopted a very weak escape criterium, and 2), in more realistic galaxy
models the density decrease in the outer parts is stronger than in
Ferrers models, and so we expect that a smaller number of stars will
be significantly affected by the CTF. Both points are addressed and
discussed elsewhere, in MC03 in a preliminary and qualitative way
for an Hernquist model at the center of a cluster, while in CM03 we
present the results of a systematic exploration of the parameter space
for untruncated galaxies both at the center and with the center of
mass in circular orbit in a spherical cluster.

In any case, the simple model here explored looks promising, with a
number of escapers in nice agreement with the observational estimates.

\begin{acknowledgements}
      We would like to thank Giuseppe Bertin and Daniel Pfenniger 
      for useful discussions and the Observatory of Geneve which has 
      let us use the GRAVITOR for our simulations. L.C. was supported 
      by the grant CoFin2000.

\end{acknowledgements}

\bibliographystyle{aa}

\end{document}